# Quantum spin excitations in a dual-core magnetic molecule

**Authors:**
Wenbin Li[1†✉], Wenwen Shi[2†], Xiaoxiao Xiao[3], Haiyan Zhu[4], Cai Cheng[5,6], Dongfei Wang[7], Lan Chen[8,9], Masahiro Haze[1], Huixia Fu[4✉], Xiao Zheng[10], Yang Guo[11✉], Zhendong Li[3], Yukio Hasegawa[1]

**Affiliations:**
[1]Institute for Solid State Physics, The University of Tokyo; Chiba 277-8581, Japan.
[2]CAS Key Laboratory of Precision and Intelligent Chemistry, University of Science and Technology of China; Hefei 230026, China.
[3]Key Laboratory of Theoretical and Computational Photochemistry, Ministry of Education, College of Chemistry, Beijing Normal University; Beijing 100875, China.
[4]Center of Quantum Materials and Devices & Chongqing Key Laboratory for Strongly Coupled Physics, College of Physics, Chongqing University; Chongqing 401331, China.
[5]School of Physics and Electronic Engineering, Sichuan Normal University; Chengdu, 610101, China.
[6]School of Materials and Energy, State Key Laboratory of Electronic Thin Film and Integrated Devices, University of Electronic Science and Technology of China; Chengdu, 610054, China.
[7]State Key Laboratory of Chips and Systems for Advanced Light Field Display; Key Laboratory of Advanced Optoelectronic Quantum Architecture and Measurement, Ministry of Education; School of Physics, Beijing Institute of Technology, Beijing 100081, China.
[8]Institute of Physics, Chinese Academy of Sciences; Beijing 100190, China.
[9]School of Physical Sciences, University of Chinese Academy of Sciences; Beijing 100190, China.
[10]Department of Chemistry, Fudan University; Shanghai 200433, China.
[11]School of Chemistry and Chemical Engineering, Shandong University; Qingdao 266237, China.
†These authors contributed equally: Wenbin Li, Wenwen Shi
✉Email: wli@issp.u-tokyo.ac.jp; hxfu@cqu.edu.cn; yang.guo@sdu.edu.cn

**Abstract:**
Magnetic excitations are important quantum phenomena in magnetic systems and have been widely studied in individual magnetic atoms and molecules as well as their assembled structures over the past few decades. Using scanning tunneling microscopy/spectroscopy (STM/S) combined with density functional theory (DFT) and the state-of-the-art ab initio wavefunction calculations, we investigated the properties of a novel dual-core $Cr_2Br_6$ molecule, which consists of two Cr ions coupled via superexchange through a single near-90° Cr-Br-Cr scissors bond. Under zero magnetic field, we observed a Fano peak with multi-steps through STS. When an external magnetic field is applied, some steps exhibit additional splitting, while others change little. We find that the $Cr_2Br_6$, exhibits a spin-degenerate ground state, and the complex peak splitting arises from the coexistence of vibrational and magnetic excitations in the molecule. Our results reveal rich quantum spin behavior in a well-defined two-core magnetic trihalide complex at the atomic scale, offering not only a minimal model for superexchange-coupled multi-spin quantum excitations but also a possible foundational unit for future molecule-based quantum functionalities.

**Introduction:**
Magnetic excitations in individual atoms and molecules are important phenomena in the field of quantum magnetism, drawing considerable attention due to their fundamental physical significance and potential applications, such as in spin qubits[1–3]. Beyond these aspects, the study of magnetic excitations in simple atomic and molecular systems also serve as important platforms for understanding quantum phenomena in more complex magnetic materials and their hybrid systems with metals or superconductors, including magnons[4], Kondo lattices[5,6], and topological superconductivity[7–9]. Owing to the atomic resolution capability of STM/S, the magnetic excitations in single atomic and molecular magnets on surfaces can be directly probed, typically appearing as a mirror-symmetric stepwise spectrum near the Fermi surface in the $dI/dV$ spectra[1,10]. Through atom manipulation or precise on-surface synthesis, the quantum behavior of assembled magnetic atoms and nanographene molecules with various sizes and configurations, along with the emergent Kondo lattice and topological quantum magnetic states which are primarily driven by the Ruderman-Kittel-Kasuya-Yosida (RKKY) or direct spin interactions[1,5,6,11–22], have been thoroughly studied, benefiting from their well-defined structure and spin models. Magnetic molecules with superexchange-coupled multi-core structures have also been studied in large molecular systems, such as magnetic ion based complexes[23], which often comprise numerous atoms with intricate geometries and diverse chemical bonds. The complexity of their structural geometries and variable valence states can significantly influence the quantum spin properties, posing major challenges in capturing strongly correlated open-shell electrons and degenerate magnetic energy levels. These difficulties often render conventional DFT inadequate due to its single-reference limitations. To overcome these limitations, high-accuracy multi-reference (MR) wave function methods are essential for accurately characterizing magnetic excitations in atoms or molecules[24]. However, due to the huge computational costs, traditional MR methods are limited to single-core magnetic molecules. To study the magnetic excitations in multi-core systems, advanced methods such as the ab initio density matrix renormalized group (DMRG) algorithm[25,26] or approximate selected configuration interaction (CI) methods have been extensively developed in the past decade[27–

[29]. These methods enable the investigation of the electronic structures of multi-nuclear systems, such as exploring the low-lying excited states of iron-sulfur clusters[30,31] and analyzing the superexchange mechanism in cuprates with corner-sharing $CuO_4$ plaquettes[32]. For such compounds, precise structures determined by experimental probing are essential for the computational consideration of spin symmetry and spin–orbit coupling effects, which are crucial for accurately describing magnetic excitations, although such experimental determinations remain highly challenging.

In this work, we prepared a new kind of $Cr_2Br_6$ individual molecules on the Au(111) substrate by evaporating single crystals of the vdW magnet $CrBr_3$. The precise molecular structure and adsorption configuration were determined by combining atomically resolved STM imaging with DFT calculations, followed by probing the quantum excitations through STS. The d$I$/d$V$ spectrum of the $Cr_2Br_6$ molecule exhibits a Fano peak with splittings, likely arising from the coexistence of the Kondo effect and excitation modes (denoted as E1, E2, E3 and E4). Magnetic-field-dependent measurements reveal additional splittings in the spectrum (denoted as E0 and E5), with their excitation energies increasing linearly with the applied field. DFT calculations demonstrated that the lowest energy levels of vibrational excitations correspond to the hindered rotational and translational vibrations of the $Cr_2Br_6$ molecule, with energies of 1.4 and 2.2 meV, respectively, explaining the origin of the E2 and E3 excitations. High-level DMRG and iterative configuration interaction (iCI) calculations suggest that the ground state of the $Cr_2Br_6$ molecule corresponds to S = 3, indicating ferromagnetic coupling between the two Cr ions. When the spin-orbit coupling (SOC) effect is considered, the septet ground state (S = 3) further splits into four energy levels with degeneracies of 2-2-2-1, with a doubly degenerate ground state. The coexistence of magnetic and vibrational excitations gives rise to the complex spectral splitting features in experiments. Our study on $Cr_2Br_6$ revealed the quantum spin behavior of a well-defined dual-core magnetic molecule by combining atomically resolved structural characterization with state-of-the-art quantum chemistry calculations, providing a model platform for exploring quantum magnetism in multi-core molecular systems.

**Results and discussion:**
After depositing a small amount of $CrBr_3$ on the Au(111) surface at room temperature, rectangular bright spots surrounded by a network of spherical spots are observed on the surface, as shown in Fig. 1a. According to the previous reports, the network of spherical spots are the Br atoms adsorbed on the Au(111) surface[33,34], which also suppresses the native herringbone reconstruction of the substrate. In contrast, the rectangular bright spots should correspond to one kind of chromium bromide molecular structure. The high-resolved STM image (Fig. 1b) indicate that the molecule has a 'double-X' like structure with brighter contrast at each 'X' center. A molecular model of $Cr_2Br_6$ is proposed, where all eight atoms lie nearly in one plane, with two Cr ions situated at the centers of 'X'-shaped motifs and six Br atoms occupying their terminal vertices (Fig. 1c). In this molecule, the two magnetic ions are connected by a near-90° Cr–Br–Cr scissors bond, which corresponds to the Cr-Cr coupling pathway in the van der Waals magnet $CrBr_3$[35]. The precise molecular adsorption sites on the Au(111) substrate were determined by comparing the Br atom configurations (Supplementary Fig. 1a and b). Based on this, we performed DFT calculations for $Cr_2Br_6$ on the Au(111) surface with and without the surrounding Br atoms. The optimized structure (Fig. 1c) reveals a stable, slightly deformed planar configuration, with the Cr atoms positioned closest to the Au surface (2.60 Å) and the Br atoms at the bridging sites farthest (3.03 Å). Correspondingly, the STM simulation (Fig. 1d and Supplementary Fig. 1e) further validates the accuracy of this molecular model.

To investigate the magnetic properties of the $Cr_2Br_6$ molecule, we conducted d$I$/d$V$ spectroscopy measurements on the molecule and observed a Fano peak at the Fermi level (as shown in Fig. 2a), which may originate from Kondo resonance. The d$I$/d$V$ spectrum measured on the Au(111) substrate, located near this molecule using the same micro-tip, shows the same tilted background observed on the molecule which may be attributed to the tip. Figure 2b shows the on-molecule d$I$/d$V$ spectrum subtracted by the off-molecule spectrum, revealing a clearer Fano-shaped curve, which can be well-fitted by the Fano Kondo function[11]:

$$\frac{dI}{dV}(V) \propto \frac{(\epsilon + q)^2}{1 + \epsilon^2}, \qquad \left(\epsilon = \frac{eV - \epsilon_0}{\Gamma}\right), \tag{1}$$

where $q$ reflects the curve shape of the Fano resonance, which depends on the tunneling junction between the tip and the sample, $\epsilon_0$ is the energy location of the resonance, and $\Gamma$ is the half-width at half-maximum of the resonance (HWHM), depends on the finite temperature $T$ and Kondo temperature $T_k$ as[11]:

$$2\Gamma = \sqrt{(\alpha k_B T)^2 + (2k_B T_k)^2} \tag{2}$$

where $\alpha$ is the constant representing the linear dependence between $T$ and $\Gamma$ in the case of $T \gg T_k$. In the fitting results in Fig. 2b, $\Gamma \approx 4.5$ meV, which corresponding a $T_k$ much higher than our measuring temperature of $T = 0.36\ K$. Thus, the first part of the $\Gamma$ in (2) can be neglected and the $T_k$ can be estimated as: $T_k = \Gamma/k_B \approx \Gamma \times 11.6\ K\ meV^{-1} = 51\ K$.

Interestingly, in the highly-resolved spectrum (Fig. 2c), mirror symmetric multi-steps are observed, likely originating from the inelastic tunneling process of the vibrational or magnetic excitations. From the numerical smoothed d$I$/d$V$ and differential (d$^2I$/d$V^2$) curves (Supplementary Fig. 2), four pairs of inelastic tunneling spectroscopy (IETS) excitation steps can be roughly identified. To determine the precise excitation energies and decompose the contributions of each component, a fitting function, incorporating a Fano peak and four IETS steps with a Fano background, is applied to the data (see details in Supplementary Note 2). The successful fitting result (Fig. 2c, d) confirms the accuracy of the function and the precise energy levels of the four excitation modes. Furthermore, the site-dependent d$I$/d$V$ measurements at different locations of the molecule (Fig. 2e) show a homogeneous spectral characteristic. Additional fitting analysis of the spectra (Supplementary Fig. 3) shows that all the IETS steps have no obvious site-dependent differences, indicating that the two Cr ions in the molecule are equivalent.

To further confirm the mechanism behind the four excitation modes, we performed magnetic-field-dependent d$I$/d$V$ measurements, which revealed a clear difference in the IETS steps under magnetic fields. The numerical second differential conductance (d$^2I$/d$V^2$) under various magnetic fields (Supplementary Fig. 4a) showed that the four excitation modes split into six modes. Notably, E4 and E5 are too close to be clearly distinguished, so they can also be treated as a single excitation mode. However, the theoretical calculations suggest that analyzing with six excitation modes is more reasonable. To decompose the contribution of each component more precisely, a fitting function that combines a Fano peak (Kondo) and six IETS (E0-E5) was applied to the data (Fig. 3a, with fitting details in Supplementary Fig. 4b-f). The fitting results indicate that the excitation energies of E1, E3, E4 and E5 ($\Delta_1$, $\Delta_3$, $\Delta_4$, $\Delta_5$) remain almost constant across various magnetic fields, while the $\Delta_2$ and $\Delta_6$ are dependent on the magnetic field, likely due to the Zeeman effect on the spin states. To further explore the magnetic field dependence of the fitted excitation energies ($\Delta$), linear functions are used for fitting:

$$E_i(B) = a_i B + c_i \tag{3}$$

where $a_i$ represent the slope, directly proportional to the $g$ factor, $B$ is the magnetic field, and $c_i$ is the $\Delta$ at zero field.

The fitting results (Fig. 3i) show that E0 and E5 exhibit clear slopes ($a_0 \approx 0.104$ mV/T, $a_5 \approx 0.087$ mV/T), while the slopes of other excitation modes are close to zero ($a_1 \approx 0.002$ meV/T, $a_2 \approx -0.023$ meV/T, $a_3 \approx 0.007$ meV/T, $a_4 \approx 0.006$ meV/T, the fitting error is about ±0.012 meV/T), indicating that their excitation energies remain nearly constant under magnetic fields. The fitted $c_i$ values ($c_1 \approx 0.23$ meV, $c_2 \approx 1.14$ meV, $c_3 \approx 1.90$ meV, $c_4 = c_5 \approx 2.39$ meV) represent the excitation energies at zero field, which align well with the results in Fig. 2c and d.

Typically, if excitation energy depends on external magnetic fields, it is spin-related. To better understand the magnetic-field-dependent and independent excitations observed in the experiment, the vibrations and spin states of the $Cr_2Br_6$ molecule were investigated through theoretical calculations, considering that each Cr ion has three unpaired electrons. Firstly, DFT calculations are performed to search for the vibration modes of the $Cr_2Br_6$ molecule on the Au(111) substrate. The results indicate that the three lowest-energy vibrations are 1.413 meV, 2.180 meV and 2.984 meV, corresponding to the hindered rotational and translational vibrations of the $Cr_2Br_6$ molecule, respectively (Supplementary Note 3). Comparing these results with experimental observations, it can be inferred that the excitation modes E2 (~1.1 meV) and E3 (~1.9 meV) (or E4 (~2.39 meV)), which have energies close to the value of 1.413 meV and 2.18 meV, should originate from the molecule vibrations. However, the E1, with an energy of approximately only 0.2 meV, cannot be explained by molecular vibrations. According to the previous reports, this energy is also far away from the phonon mode of the Au(111) surface[36,37]. Therefore, the most plausible explanation is that E1 also originates from spin excitation.

Subsequently, to determine the spin ground state of $Cr_2Br_6$, state-of-the-art quantum chemistry methods, namely ab initio DMRG and iCI, were employed based on the optimized molecule structure (Supplementary Note 4). Calculations were carried out using an active space of CAS(26,20), with detailed computational settings provided in the Supplementary Fig. 6. Both DMRG and iCI methods consistently predicted that the septet state (S = 3) is the ground state (Table 1), where the three singly occupied 3d electrons on each Cr(III) are ferromagnetically coupled. When SOC is included, the spin states have additional splitting. Specifically, the ground septet state splits into four energy levels with degeneracies of 2-2-2-1 (as shown in Table 2). The ground state (spinor state 1 and 2) and first excited state (spinor state 3 and 4) are both clearly identified as doubly degenerate states. When an external magnetic field is applied, the doubly degenerate ground state and excited states will experience additional splittings due to the Zeeman effect.

Comparing experimental observations with theoretical calculations, a schematic of the spin states and vibrations is proposed to explain the excitation modes observed in the $Cr_2Br_6$ molecule (as shown in Fig. 3j). In this schematic, E2 and E3 are attributed to the excitations of the molecule's rotational and translational vibrations, respectively, while the other excitation modes originate from quantum spin excitations. At zero field, the lowest six spin states exhibit three energy levels with degeneracies of 2-2-2. E1 and E4 correspond to excitations from the doubly degenerate ground state (spinor state 1,2) to the doubly degenerate first (spinor state 3,4) and second (spinor state 5,6) energy levels, respectively. When an external magnetic field is applied, all the doubly degenerate spinor states undergo further Zeeman splitting into non-degenerate states (labeled 1, 3, 2, 4, 5 and 6). The excitation energies (E1 and E4 under magnetic fields) from the ground state (state 1) to the excited majority states (states 3 and 5, marked as blue line in Fig. 3j) would change little under magnetic fields. In contrast, the excitations (E0 and E5) from the ground state to the excited minority states (states 2 and 6) exhibit an obvious increase in their energies linearly dependent on the external magnetic field. In principle, there should be another excitation from the ground state (state 1) to state 4, but it has been neglected in this analysis and included in E0, as the energy difference between state 2 and state 4 is too small. It is worth noting that the energy differences between the ground and excited states, as predicted by the SOiCI calculations, are significantly larger than those observed in the experiments. One possible explanation for this deviation is that the energy resolution of the calculation method does reach the meV scale, although the results of degeneracies are unaffected. Another possibility is that the Au(111) substrate would induce an additional effect on the magnetic energy eigenlevels but have not been included in our quantum chemistry calculations, whereas previous studies suggested that charge transfer from the substrate would decrease the excitation energy of the adsorbed molecule[38] and the study on $CrBr_3$ single layer have also shown that the energy band can be compressed by the substrate[39].

By combining magnetic-field-dependent STM/S measurements with theoretical calculations, we identified the quantum vibrational and magnetic excitations induced inelastic tunneling processes in the $Cr_2Br_6$ molecule. The calculations revealed a septet ferromagnetic coupled spin ground state, which splits into 2-2-2-1 degeneracies after SOC is included, providing an explanation for the experimental observations. Our study offers valuable insight for understanding the quantum magnetic behavior in a well-defined magnetic molecule with a single transition-metal-halogen-transition-metal scissors bond, and provides a representative example for exploring complex quantum phenomena in superexchange coupled multi-core magnetic systems through atomically resolved experimental characterization combined with high-level ab initio wavefunction calculations with large active spaces.

**Methods:**

**Sample:** The sample was prepared using the molecular beam epitaxy (MBE) method within an ultra-high vacuum (UHV) environment with base pressure of about $2 \times 10^{-8}$ Pa. The clean Au(111) surface was prepared through two cycles of 20-minute sputtering (1kV, Ar pressure $3 \times 10^{-3}$ Pa) followed by 20-minute annealing at 500 °C. The $Cr_2Br_6$ molecules were deposited onto the clean Au(111) surface at room temperature by evaporating the $CrBr_3$ single crystal bought from HQ graphene.

**STM/S measurements:** The in-situ STM/S measurements were performed in a Unisoku 1300 system equipped with a superconductor magnet capable of applying an out-of-plane magnetic field up to 7 T. All STM measurements were performed at 360 mK using a PtIr tip. The STM topography images and d$I$/d$V$ spectra were analyzed using the WSxM software[40] and Python code.

**Density functional theory calculations:** For the analysis of structural configurations and vibrational modes of the $Cr_2Br_6$ molecule on the Au(111) surface, we used the Vienna Ab-initio Simulation Package (VASP)[41,42], a widely utilized tool for density functional theory (DFT) calculations that incorporates plane-wave pseudopotentials and the projector-augmented wave (PAW) method. The Au(111) surface was modeled using a three-layer slab in an 8×5 supercell configuration with a 17 Å vacuum layer added along the z-axis to eliminate interactions from the periodic boundary conditions. Geometric optimization was performed by fixing the bottom Au layer, allowing all other atoms relaxed fully. The exchange-correlation energy was calculated using the generalized gradient approximation with the Perdew-Burke-Ernzerhof (GGA-PBE) functional with Grimme's DFT-D3 dispersion corrections to accurately account for van der Waals interactions[43,44]. An energy cutoff of 400 eV was applied, with convergence thresholds set at $1 \times 10^{-4}$ eV/atom for energy and $1 \times 10^{-2}$ eV/Å for atomic forces. Due to the large system size, Γ-point sampling was applied to accelerate calculations. Simulated STM images were generated using P4VASP software for comparison with experimental data.

**Ab initio wavefunction calculations (DMRG and iCI):** In all quantum chemistry calculations, the TZP-DKH basis sets were employed [45]. To account for scalar relativistic effects, the spin-free exact two-component (sf-X2C) Hamiltonian was used [46–48]. To include the spin-orbit coupling effect, the first-order Douglas-Kroll-Hess mean-field one-body spin-orbit Hamiltonian (so-DKH1) was utilized [49]. The geometry of $Cr_2Br_6$ was optimized at the UTPSSh-D3BJ level. Based on this optimized geometry, the electronic structure of $Cr_2Br_6$ was computed using CASCI(26,20) [26 active electrons in 20 active orbitals] via the ab initio DMRG using the StackBlock code [50] and a selected CI method, iCI [29,51], implemented in the BDF package. The SOiCI and SOiCI(2) [with second-order correction] methods were used to further takes into account the spin-orbit couplings based on the iCI calculations.

Correlated Electrons. *J. Chem. Theory Comput.* **17**, 949–964 (2021).

## Acknowledgement:

This work was partially supported by the Grants-in-Aid for Scientific Research (KAKENHI) from the Japan Society for the Promotion of Science (Grants No. JP19H00859, No. JP20K15166, No. JP22K14598, No. JP22H00292, No. JP23KF0136). W. Li acknowledges the support by the JSPS postdoc fellowship (No. P23054). Y. Guo acknowledge the support by National Natural Science Foundation of China (Grant No. 22273052, 22433001). X. Xiao and Z. Li acknowledge the support from the Innovation Program for Quantum Science and Technology (Grant No. 2023ZD0300200) and the Fundamental Research Funds for the Central Universities. H. Fu acknowledges the support from the National Natural Science Foundation of China (Grants No. 12104072, No. 12347101).

## Author contributions:

W. L. proposed the research project. W. L. performed the experiments and analyzed the data in discussion with D. W., L. C., M. H., and Y. H.. W. S., H. Z., C. C., performed the DFT calculations under the supervision of H. F., and X. Z.. X. X., Y. G., and Z. L. performed the DMRG and iCI calculations. W. L. wrote the first version of the manuscript and finished the paper with input from all other authors. The paper reflects the contributions of all authors.

## Competing interests:

The authors declare no competing interests.

## Materials & Correspondence

Correspondence and requests for materials should be addressed to Wenbin Li, Huixia Fu and Yang Guo.

**Figures and tables:**

**Fig. 1**

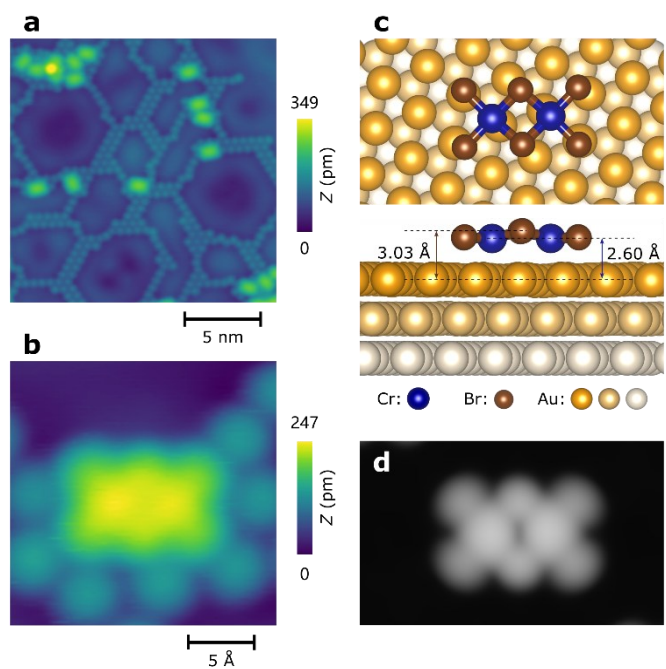

**Fig. 1 | Structure of the Cr$_2$Br$_6$ molecule. a,** STM topography image of sample surface after deposition of a small amount of CrBr$_3$ on Au(111) surface. The brighter rectangle spots represent the Cr$_2$Br$_6$ molecules, while the surrounding smaller spots organized network are self-assembled Br atoms on the Au(111) surface. Setpoint: $V_{sample}$ = 100 mV, $I$ = 20 pA. **b,** High-resolved image of single Cr$_2$Br$_6$ molecule, surrounded by some Br atoms. Setpoint: $V_{sample}$ = 10 mV, $I$ = 800 pA. **c,** Schematic illustration showing the relationship between the Cr$_2$Br$_6$ and CrBr$_3$ layer. **d,** Structure model of the Cr$_2$Br$_6$ on Au(111) surface, optimized by DFT calculations. **e,** STM simulation of the Cr$_2$Br$_6$ molecule, corresponding to the experimental results shown in **b**.

**Fig. 2.**

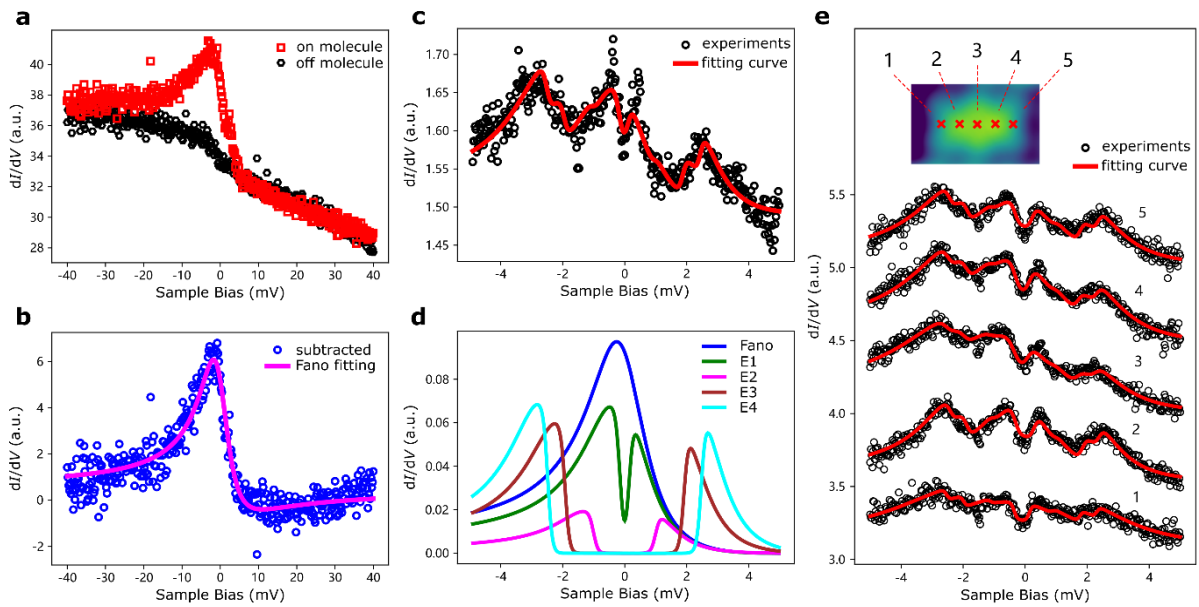

**Fig. 2 | d$I$/d$V$ spectra on Cr$_2$Br$_6$. a,** d$I$/d$V$ spectrum measured on the top of the Cr$_2$Br$_6$ molecule (red rectangle) and on the nearby Au(111) substrate (off molecule, black hexagonal) with the same tip. Setpoints: $V_{sample}$ = 50 mV, $I$ = 2 nA, $f_{lock-in}$ = 971 Hz, $V_{lock-in}$ = 1 mV. **b,** The d$I$/d$V$ spectrum on-molecule subtracted by the one off-molecule (blue circle) in **a** and the fitting results (magenta curve) with a Fano function. The fitted $\Gamma$ = 4.4 meV, corresponds to a Kondo temperature of $T_k$ ≈ 51 K. **c,** High-resolved d$I$/d$V$ spectrum (black circle) in a small energy-scale and the fitting results (red curve) with the sum function of a Fano background and four ITES steps. Setpoints: $V_{sample}$ = 50 mV, $I$ = 2 nA, $f_{lock-in}$ = 971 Hz, $V_{lock-in}$ = 50 μV. **d,** The fitted Fano and IETS components of the red curve in **c**. According to the fitting results, the IETS excitation energies of the four excited states (E1, E2, E3, E4) are: $\Delta_1$ ≈ 0.14 meV, $\Delta_2$ ≈ 1.0 meV, $\Delta_3$ ≈ 1.9 meV, $\Delta_4$ ≈ 2.5 meV. **e,** Site-dependent d$I$/d$V$ spectra (black circle) and their corresponding fitting curve (red curve) on the Cr$_2$Br$_6$ molecule, showing uniform spectral characteristics. Site-2 and site-4 are located atop Cr atoms. Setpoints: $V_{sample}$ = 50 mV, $I$ = 2 nA, $f_{lock-in}$ = 971 Hz, $V_{lock-in}$ = 100 μV. The inset image shows the sites where the spectra were measured. The fitting details are shown in Supplementary Fig. 2.

**Fig. 3.**

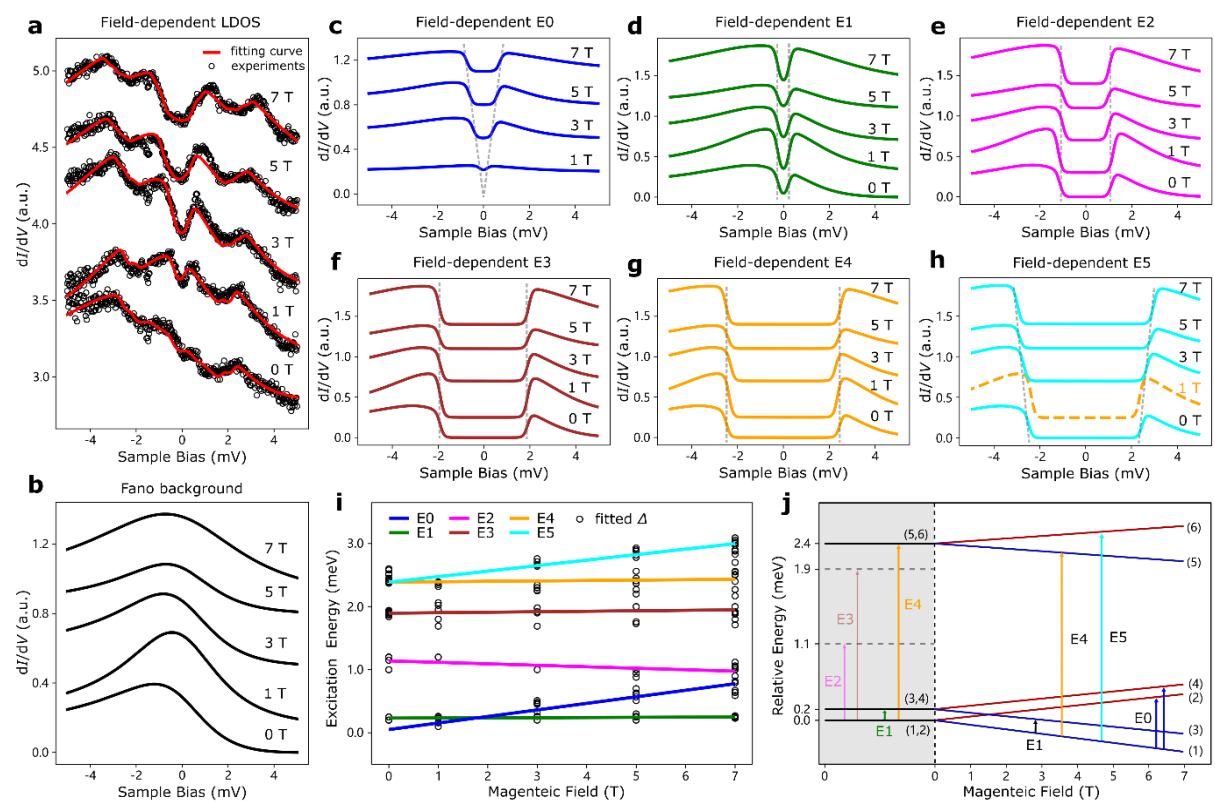

**Fig. 3 | Magnetic-field-dependent d$I$/d$V$ spectra on $Cr_2Br_6$. a,** The d$I$/d$V$ spectra measured on the same $Cr_2Br_6$ molecule under various magnetic fields. The experimental data (black circles) and corresponding fitting curves (red line) are shown. Setpoints: $V_{sample}$ = 50 mV, $I$ = 2 nA, $V_{lock-in}$ = 100 μV, $f_{lock-in}$ = 971 Hz. The experimental spectra are fitted by a sum function combining a Fano background and six IETS steps (see Supplementary Fig. 2 for details). **b-h,** Normalized fitting components at different magnetic fields. E1, E2, E3 and E4 remain mostly constant with varying magnetic fields, while excitation energies of E0 and E5 increase as the magnetic field increases. Notably, excitation energy of E0 is zero in zero filed, indicating it originates from the splitting of a two-fold degenerate spin state. The dashed curve in **h** corresponds to the 1 T curve from **g**. **i,** Statistical analysis of the fitted excitation energies (E0 - E5) as a function of magnetic field. The field-dependent excitation energies ($\Delta$) are fitted with a linear function $E_i(B) = a_iB + c_i$. Fitting results for each excitation mode are: E0: a0 ≈ 0.104 meV/T, c0 ≈ 0.05 meV; E1: a1 ≈ 0.002 meV/T, c1 ≈ 0.23 meV; E2: a2 ≈ -0.023 meV/T, c2 ≈ 1.14 meV; E3: a3 ≈ 0.007 meV/T, c3 ≈ 1.90 meV; E4: a4 ≈ 0.006 meV/T, c4 ≈ 2.39 meV; E5: a5 ≈ 0.087 meV/T, c5 = c4. The fitting errors are about ±0.012 meV/T for $a_i$, and about ±0.05 meV for $c_i$. **j,** Schematic representation of the possible spin states (blue and red lines).

**Table 1:**

| S | DMRG-CASCI (M = 5000) | | iCI-CASCI (Cmin = 7.5 × 10$^{-6}$) | |
|---|---|---|---|---|
| | Absolute Energy (Hartree) | Relative Energy (meV) | Absolute Energy (Hartree) | Relative Energy (meV) |
| 0 | -17725.48526 | 27.4 | -17725.48527 | 27.6 |
| 1 | -17725.48541 | 23.3 | -17725.48542 | 23.7 |
| 2 | -17725.48573 | 14.5 | -17725.48574 | 14.9 |
| 3 | -17725.48627 | 0 | -17725.48629 | 0 |
| 4 | -17725.39110 | 2589.7 | -17725.39112 | 2589.7 |

**Table 1.** The spin energies of $Cr_2Br_6$ computed by spin-adapted DMRG-CASCI using bond dimension M=5000 and iCI-CASCI with CAS(26,20) active space.

**Table 2:**

| state | SOiCI (Hartree) | relative energy (meV) | SOiCI(2) (Hartree) | relative energy (meV) |
|---|---|---|---|---|
| 1 | -17725.48690 | 0.00 | -17725.48710 | 0 |
| 2 | -17725.48690 | 0.00 | -17725.48710 | 0 |
| 3 | -17725.48658 | 8.74 | -17725.48679 | 8.5 |
| 4 | -17725.48658 | 8.75 | -17725.48679 | 8.6 |
| 5 | -17725.48640 | 13.61 | -17725.48662 | 13 |
| 6 | -17725.48639 | 13.84 | -17725.48660 | 13.5 |
| 7 | -17725.48636 | 14.76 | -17725.48658 | 14.1 |

**Table 2**. The lowest seven absolute and relative energies of $Cr_2Br_6$ computed by SOiCI-CASCI and SOiCI(2)-CASCI with CAS(26,20) active space with SOC are included.